\documentclass[12pt,letter]{article}

\usepackage[latin1]{inputenc}
\usepackage{graphicx} 
\usepackage{amsmath}
\usepackage{amsfonts}
\usepackage{amssymb}
\usepackage{geometry}
\usepackage{listings}
\title{MODA: MOdule Differential Analysis for weighted gene co-expression network}
\author{\textsc{Dong Li}
\\{School of Computer Science, The University of Birmingham, UK}\\
\textsc{James B. Brown}
\\{Department of Statistics, University of California Berkeley, USA}\\
\textsc{Luisa Orsini}
\\{School of Biosciences, The University of Birmingham, UK}\\
\textsc{Zhisong Pan},\textsc{Guyu Hu}
\\{PLA University of Science and Technology, China}\\
\textsc{Shan He}
\\{School of Computer Science, The University of Birmingham, UK}\\
} 
\begin{document}


\maketitle
\section{Summary}
Gene co-expression network differential analysis is designed to help biologists understand gene expression patterns under different condition. By comparing different gene co-expression networks we may find conserved part as well as condition specific set of genes. Taking the network as a collection as modules, we use a sample-saving method to construct condition-specific gene co-expression network, and identify differentially expressed subnetworks as conserved or condition specific modules which may be associated with biological processes. We have implemented the method as an R package which establishes a pipeline from expression profile to biological explanations. The usefulness of the method is also demonstrated by synthetic data as well as Daphnia magna gene expression data under different environmental stresses.\\
\textbf{Availability:} Available at https://www.cs.bham.ac.uk/~szh/software.xhtml\\
\textbf{Contact:} {s.he@cs.bham.ac.uk}
\section{Introduction}
Gene co-expression network attracts much attention nowadays. In such a network, nodes represent genes and each edge connecting two genes stands for how much degree may this pair of genes are co-expressed across several samples. The presence of these edges is commonly based on the correlation coefficients between each gene pairs. The higher of correlation between a pair of genes, the higher probability that there exists a co-functionality relationship between them. With proper choice of minimal correlation value as a threshold, we can generate an unweighted and undirected network for given gene expression profile. But the optimal cut-off threshold is difficult to determine. And throwing away relatively large proportion of correlation coefficients will lead to information loss. In contrast,  weighted correlation network analysis (WGCNA) overcomes this drawback by keeping all possible edges but shows how significant is the co-expression relationship using edge weights \cite{langfelder2008wgcna,zhang2005general}.

A module in a biological network is defined as a subnetwork which may involves a common function in biological processes. The module detection in WGCNA is based on hierarchical clustering, which groups similar genes into one cluster. The similarity was defined by topological overlap measure \cite{zhang2005general}. Following the logic of WGCNA, here we mainly improve it from the following three aspects: 1) How to determine the cutting height of hierarchical clustering tree roughly depends on self-definition in WGCNA. Here we give an option to choose the height based on the quality of partition. 2) Edge weights in gene co-expression networks are defined by correlation coefficients of gene pairs. And it is well known that the accurate correlation coefficient is approximated by $1/sqrt(n)$ where $n$ is the number of samples, which makes it impossible to get reliable correlation coefficients with only several replicates under each experimental condition in practice. We use a sample-saving way to analyze condition-specific co-expression network for each single condition. 3) Taking a network as a collection of modules, we generalize the differential analysis from individual genes to modules, which may find condition specific and conserved subnetworks.
\section{Methods}
Inspired by the concept of partition density of link communities \cite{ahn2010link} where the modules were defined based on the link similarity, we propose a cutting method to make the average density of resulting modules to be maximal. Here we simply define the module density as the average edge weights in one module (equation (1) in supplementary file) which keeps the same in \cite{zhang2005general}, and then find the cutting height of hierarchical clustering that leads to maximal average density. We also provide other criterion such as average modularity for weighted network \cite{newman2004analysis} of resulting clusters to determine the cutting height.

General gene differential analysis has covered identification of important individual genes which shows significant changes across multiple conditions \cite{smyth2005limma}. However, based on the fact that genes interact with each other to exert some biological function instead of acting alone, it is more informative to identify a subnetwork (module) of genes which are conserved across multiple conditions or just active in certain conditions. DICER \cite{amar2013dissection} also goes beyond individual gene differential analysis, using a probabilistic framework to detect differentially co-expressed gene sets. DINA \cite{gambardella2013differential} can identify condition-specific modules from a collection of condition-specific gene expression profiles which differs from our sample-saving method. Based on a set of condition-specific networks, we use WGCNA to identify modules for different networks. Then, we use the Jaccard index, which essentially measures the similarity between two sets of elements, to measure the similarity between modules from two different networks.

By comparing all module pairs of two networks, we can get a similarity matrix $A$,where each entry $A_{ij}$ means the Jaccard similarity coefficient between the $i$-th module from the network $N_1$ and $j$-th module from the network $N_2$. Assume the $N_1$ is background, normally containing samples from all conditions, and the $N_2$ is constructed from all samples minus samples belong to certain condition $D$ \cite{kuijjer2015estimating}. Then the elements in row sum of $A$ (vector denoted by ${\bf s}$) indicate how much degree that modules in $N_1$ can be affected by condition $D$. The higher ${\bf s}_i$ means the module $i$ in $N_1$ may just be responsible for general stress. Especially when some ${\bf s}_i$  in $N_1$ keeps relatively high row sum of $A$ compared with all other $N_2$ (remove one condition each time), showing these modules have little association with any specific conditions. While lower ${\bf s}_i$ means module $i$ in $N_1$ is very different from the modules in $N_2$, which may indicate the module has some connection with condition $D$. The rationale behind this simple  criteria is based on the mechanism of correlation, i.e. which samples can make impact on the correlation coefficient while others may not? More details can be found in supplementary file part 1.

After determine which module may be condition specific, we can associate biological process with module by functional annotation enrichment analysis. The input can be gene list from the module, or overlapping just part much with others. Here we use DAVID \cite{huang2008systematic} to conduct integrative functional annotation enrichment analysis of gene list based on an R Webservice interface \cite{fresno2013rdavidwebservice}. We implemented a module differential analysis pipeline, from gene expression profile of multiple conditions to enrichment analysis results. Figure shows the general process of each step mentioned above.
\begin{figure}[h]
\centering
\includegraphics[natwidth=833,natheight=638,width=0.5\textwidth]{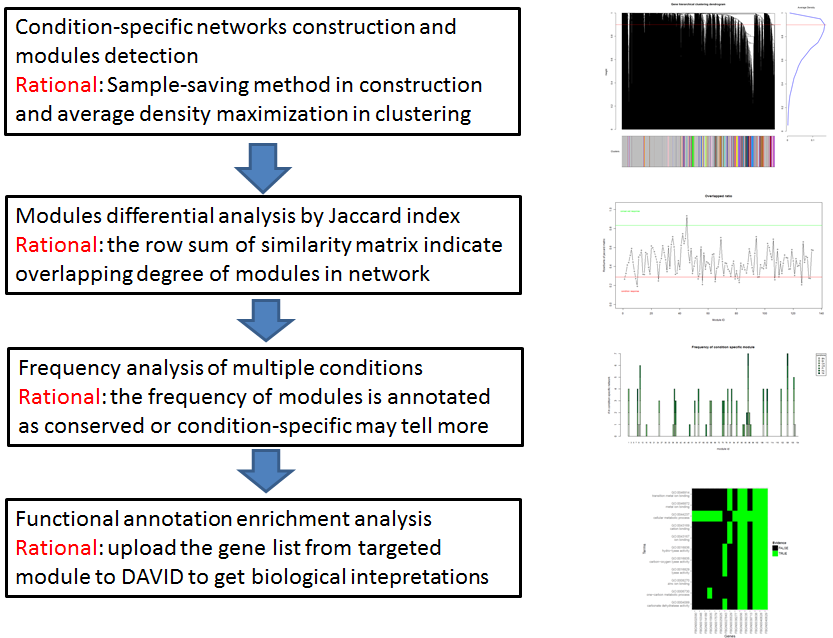}
\caption{Overview of MODA.}
\end{figure}
\section{Result}
We evaluated the effectiveness of proposed methods on both synthetic data and real-world data. By comparing two gene expression profiles generated by different desired correlation matrices of the same set of genes, we can determine the genes affected by a groups definition, which is consistent with the generator. The details for simulation as well as the usage of package can be found in supplementary file part 2. The method is also used on a comprehensive RNA-Seq data set obtained from two natural genotypes fo D. magna, to detect condition-specific as well as conserved responsive genes and biological functions. Several biological meaningful results show the capability of the algorithm, and more details can be found in [stressflea draft].
\section{Supplementary}
\subsection{Concept part}
\begin{figure}[htp!]
\centering
\includegraphics[natwidth=4096,natheight=3072,width=400bp]{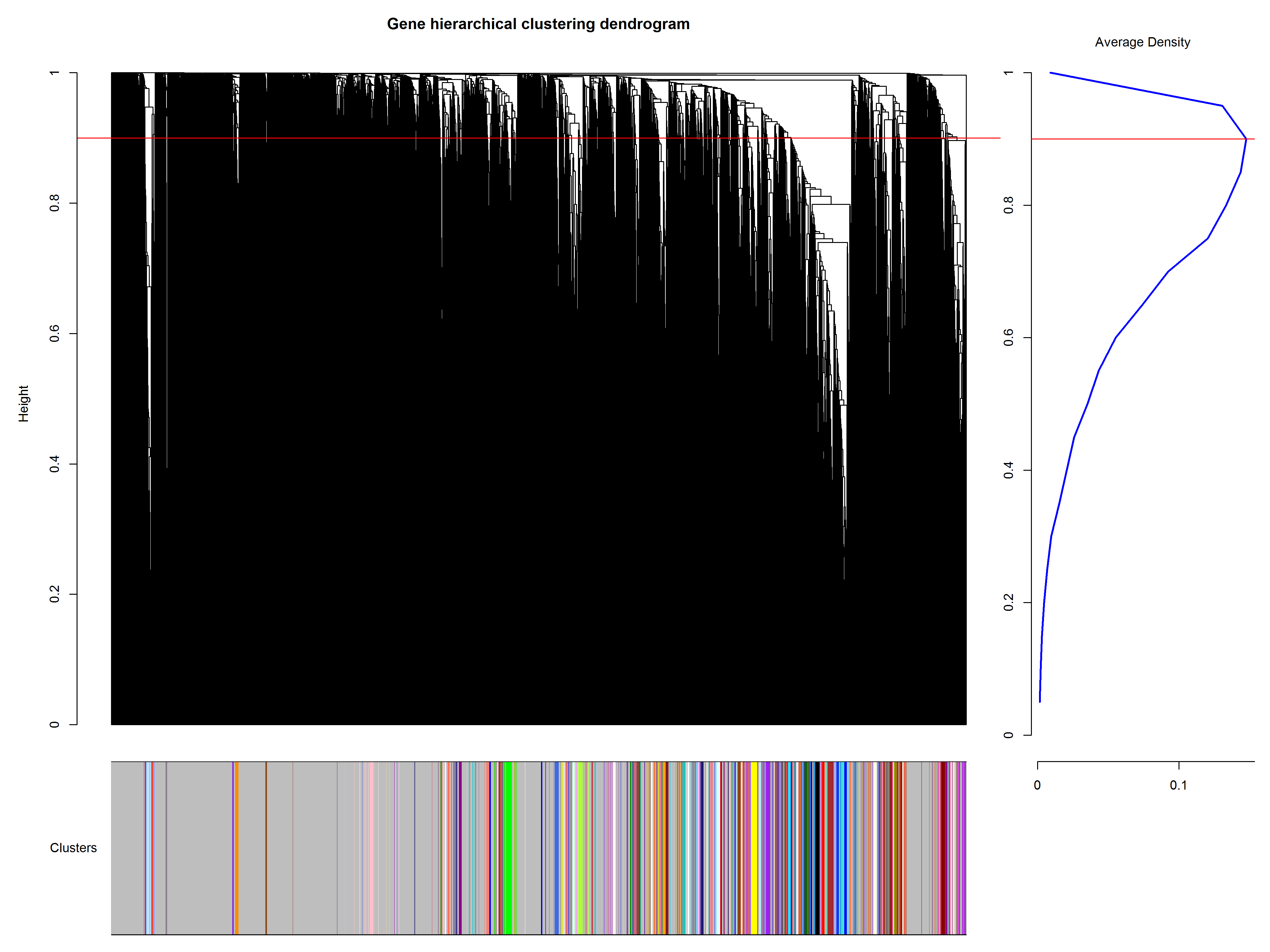}
\caption{Maximal partition density based hierarchical clustering}
\label{fig1}
\end{figure}

Given gene expression profile $X \in \mathbb{R}^{n\times p}$, where $n$ is the number of experimental samples and $p$ is the number of genes. $X_{ij}$ means the expression value of the $j$-th gene in $i$-th sample. The popular tool WGCNA \cite{langfelder2008wgcna} conducts the module detection by hierarchical clustering, i.e. putting similar gene together. The definition of similarity ranges from basic correlation to more complex topological overlap measure \cite{zhang2005general}. While how to determine the cutting height of hierarchical clustering tree remains an open problem. Here we give the option to chose the height based on the quality of partition. Inspired by the concept of partition density of link communities \cite{ahn2010link,kalinka2011linkcomm}, we choose the cutting height to make the average density of resulting modules to be optimal. The density of one module $A$ is defined as:
\begin{equation}
Density(A) = \frac{\sum_{i\in A}\sum_{j\in A,j\neq i} a_{ij}}{n_A(n_A-1)}
\end{equation}
where $a_{ij}$ is the similarity between gene $i$ and gene $j$, and $n_A$ is the number of genes in $A$. We can also use the modularity $Q$ of weighted network $A$ \cite{newman2004analysis} as the criterion to pick the height of hierarchical clustering tree:
\begin{equation}
Q = \frac{1}{2m}\sum_{ij}[a_{ij}-\frac{k_ik_j}{2m}]\sigma(c_i,c_j)
\end{equation}
where $m$ is the number of edges and $k_i$ is the connectivity (degree) of gene $i$, defined as $\sum_j a_{ij}$. And $\sigma(c_i,c_j)=1$ only when gene $i$ and $j$ are in the same module. The complete module detection and average density is shown in Figure \ref{fig1}.

\begin{figure}[htp!]
\centering
\includegraphics[natwidth=600,natheight=600,width=300bp]{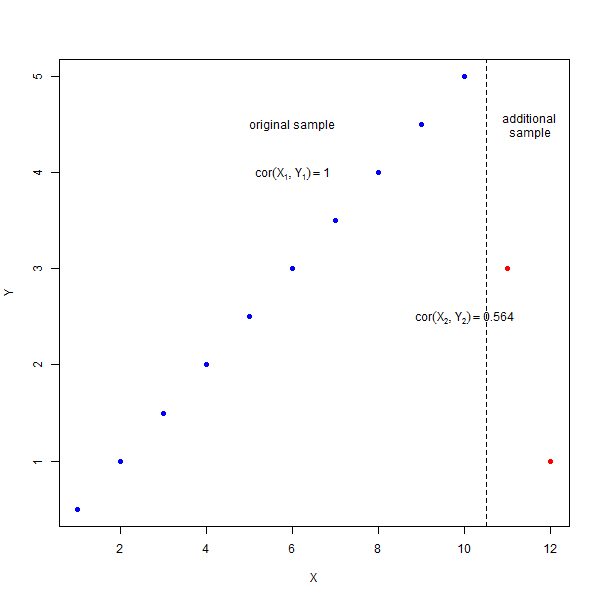}
\caption{Scatter plot of varibale $X$ and $Y$}
\label{fig2}
\end{figure}

After the module detection, the co-expression network is represented as a collection of modules (see Figure \ref{fig2}), which makes the differential analysis more focused on the modules other than the nodes or links. By comparing all module pairs from $N_1$ and $N_2$, we can get a similarity matrix $B$, where each entry $B_{ij}$ means the similarity between the $i$-th module from the network $N_1$ (denoted by $N_1(A_i)$) and $j$-th module from the network $N_2$ (denoted by $N_2(A_j)$). The similarity is evaluated by the Jaccard index.
\begin{equation}
B_{ij} = \frac{N_1(A_i)\cap N_2(A_j)}{N_1(A_i)\cup N_2(A_j)}
\end{equation}

Assume $N_1$ is background, normally containing samples from all conditions, and the $N_2$ is constructed from all samples except samples belonging to certain condition {\itshape D}. Let ${\bf s}$ is the sums of rows in $B$, i.e. ${\bf s}_i=\sum_jB_{ij}$. The value of ${\bf s}_i$ indicates how much the $i$-th module from network $N_1$ might be affected by condition {\itshape D}. The rationale behind this statistics is based on the mechanism of correlation, i.e. which samples could make an  impact on the correlation while others may not? Figure \ref{fig2} illustrates an extreme example about how the additional two samples may affect the correlation between $X$ and $Y$.

\begin{figure}[htp!]
\centering
\includegraphics[natwidth=1000,natheight=600,width=350bp]{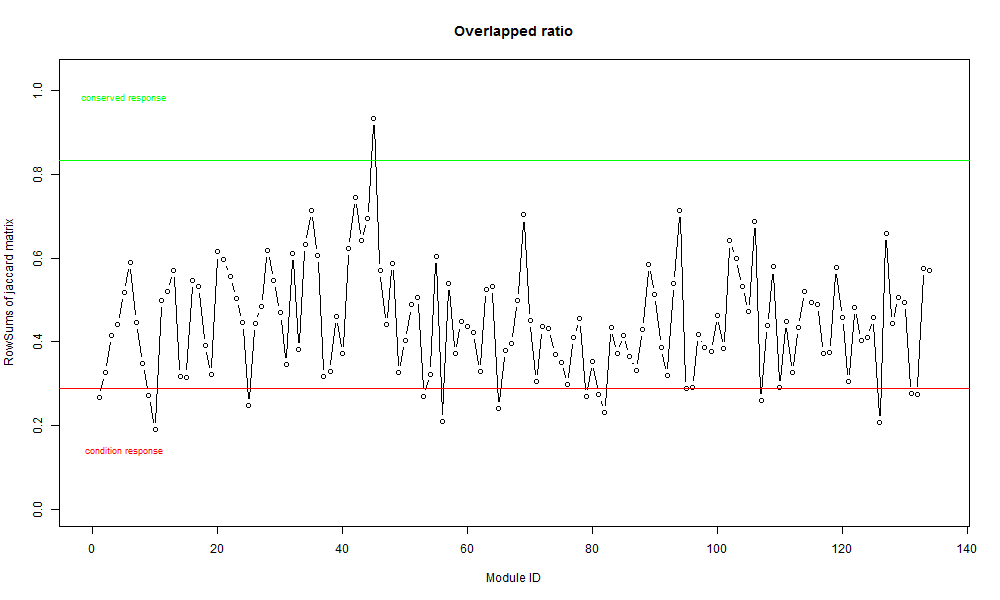}
\caption{Overlap degree of modules in $N_1$ with $N_2$}
\label{fig3}
\end{figure}

As Figure \ref{fig3} shows, we use two threshold values here: $\theta_1$ is the threshold to define $min({\bf s})+\theta_1$, less than which is considered as condition specific module. $\theta_2$ is the threshold to define $max({\bf s})-\theta_2$, greater than which is considered as condition conserved module.

We also calculate the frequency of each module is annotated as conserved or condition specific and compare all the conditions together. The rationale behind this statistics is based on the mechanism of correlation, i.e. which samples could make an impact on the correlation while others may not? The package visualizes it with a bar plot as Figure \ref{fig3}. A similar plot about the conserved module is also available. The module id is stored as a plain text file for functional enrichment analysis. Here we send one module as gene list to DAVID \cite{huang2008systematic,fresno2013rdavidwebservice} for integrative analysis.
\begin{figure}[htp!]
\centering
\includegraphics[natwidth=1000,natheight=500,width=400bp]{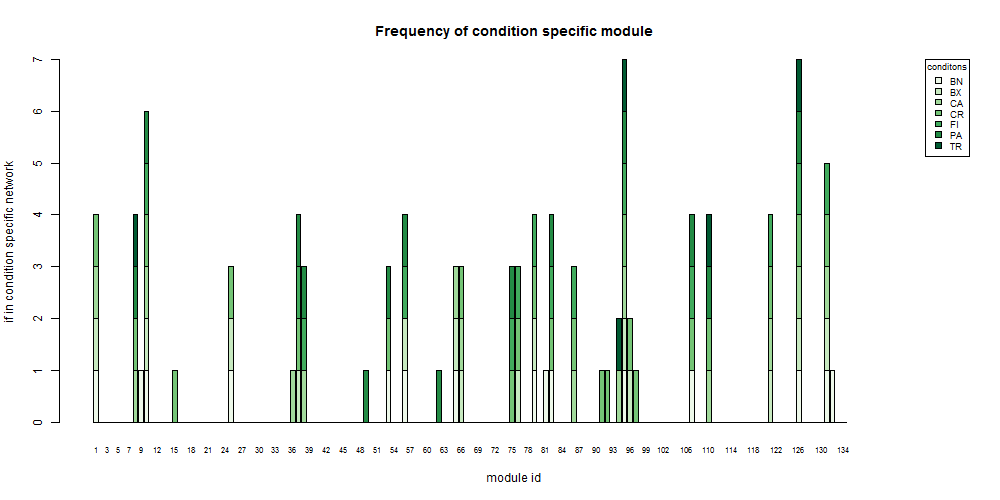}
\caption{Statistics about which module can be condition specific}
\label{fig4}
\end{figure}

\subsection{Evaluation}
\begin{figure}[htp!]
\centering
\begin{minipage}[b]{0.45\textwidth}
\includegraphics[natwidth=800,natheight=800,width=200bp]{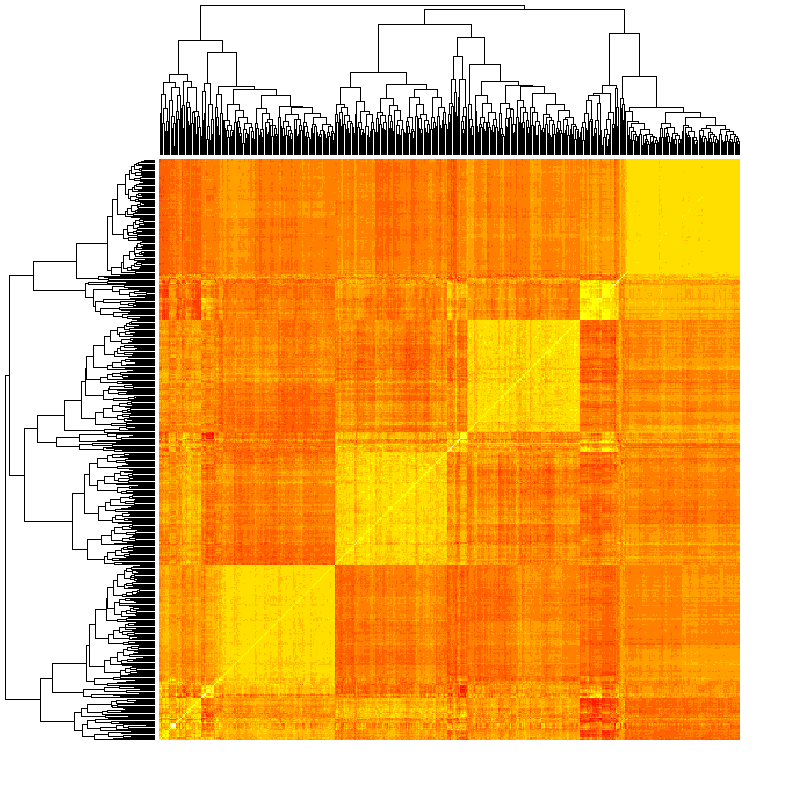}
\caption{Correlation matrix of $X$}
\label{fig5}
\end{minipage}
\begin{minipage}[b]{0.45\textwidth}
\includegraphics[natwidth=800,natheight=800,width=200bp]{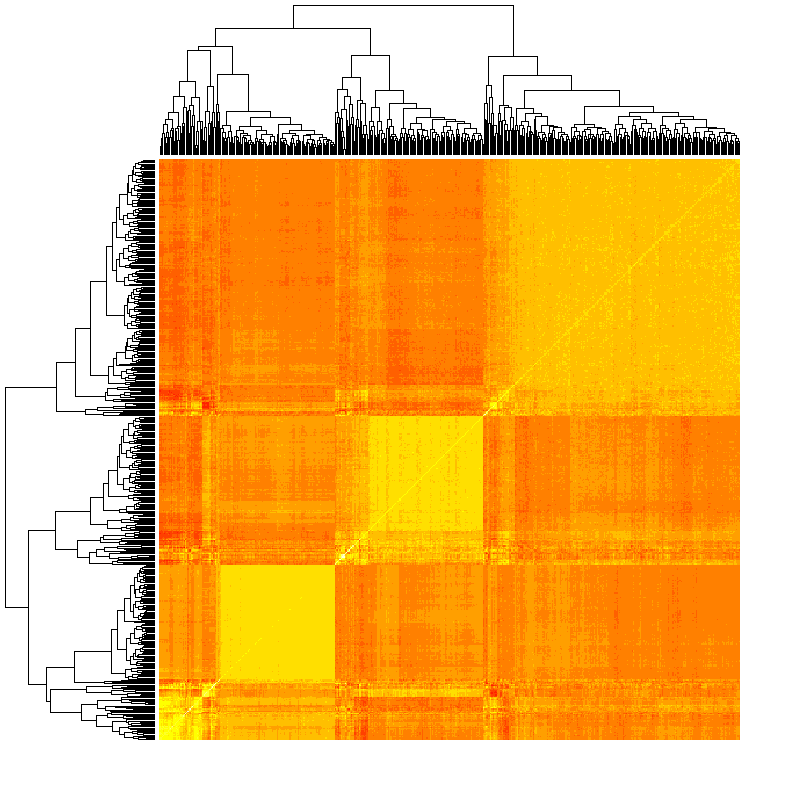}
\caption{Correlation matrix of $Y$}
\label{fig6}
\end{minipage}
\end{figure}

We evaluate the effectiveness of proposed methods on both synthetic data and real-world data. The basic synthetic gene expression data is generated by the following logic: given desired correlation matrix $C \in \mathbb{R}^{n\times p}$ with $p$ genes which has a clear modular structure that all genes are equally divided into 5 groups according to the similarities. Then we conduct the Cholesky decomposition on $C$ such that $C=LL^T$, where $L$ is the lower triangular matrix. Finally we project $L$ on random matrix $A \in \mathbb{R}^{n\times p}$ to get desired gene expression matrix $X \in \mathbb{R}^{n\times p}$, which has the rough modular structure defined by correlation $C$. Let $n=500$ and each group has 100 genes in the simulation. In each group, we allocate the gene id from 1-100, 101-200, 201-300, 301-400 and 401-500 respectively. The correlation matrix of genes in $X$ is shown in Figure \ref{fig5}. In another matrix $Y$, we merge the last two groups into one by adding more samples to $X$, and the correlation matrix is shown in Figure \ref{fig6}. The we can compare these two networks with proposed method to see which genes were affected. Gene lists in target fold show that modules that contain gene id from 1-100, 101-200 and 201-300 have large overlap with network 2, while module gene id from 301-500 which were merged have least overlap with network 2. The facts are consistent with experimental settings. 

Here is the example code to use \texttt{WeGonda} given two gene expression profiles. Results of  modules are stored under the newly created folder $\mathit{ResultFolder}$ as gene lists. The condition-specific and conserved module ids are stored as plain texts in next directory with the name of indicator which need to be compared. Other materials such as figure \ref{fig1} and \ref{fig3} are also available in the folder.
\begin{lstlisting}[language=R,basicstyle=\tiny]
library('MODA')
ResultFolder = 'ForSynthetic' # where middle files are stored
CuttingCriterion = 'Density' #CuttingCriterion could be Density or Modularity
speciesName1 = 'X' 	# indicator for data profile 1
speciesName2 = 'Y' 	# indicator for data profile 2
specificTheta = 0.1 #threshold to define condition specific modules
conservedTheta = 0.1#threshold to define conserved modules

# modules for network 1
intModules1 <- WeightedModulePartitionDensity(datExpr1,ResultFolder,speciesName1,CuttingCriterion)
# modules for network 2
intModules2 <- WeightedModulePartitionDensity(datExpr2,ResultFolder,speciesName2,CuttingCriterion)
# compare these two networks
CompareAllNets(ResultFolder,intModules,speciesName1,intModules2,speciesName2,specificTheta,conservedTheta)
\end{lstlisting}
\bibliographystyle{unsrt}
\bibliography{reference}

\begin{thebibliography}{10}

\bibitem{langfelder2008wgcna}
Peter Langfelder and Steve Horvath.
\newblock Wgcna: an r package for weighted correlation network analysis.
\newblock {\em BMC bioinformatics}, 9(1):559, 2008.

\bibitem{zhang2005general}
Bin Zhang and Steve Horvath.
\newblock A general framework for weighted gene co-expression network analysis.
\newblock {\em Statistical applications in genetics and molecular biology},
  4(1), 2005.

\bibitem{ahn2010link}
Yong-Yeol Ahn, James~P Bagrow, and Sune Lehmann.
\newblock Link communities reveal multiscale complexity in networks.
\newblock {\em Nature}, 466(7307):761--764, 2010.

\bibitem{newman2004analysis}
Mark~EJ Newman.
\newblock Analysis of weighted networks.
\newblock {\em Physical Review E}, 70(5):056131, 2004.

\bibitem{smyth2005limma}
Gordon~K Smyth.
\newblock Limma: linear models for microarray data.
\newblock In {\em Bioinformatics and computational biology solutions using R
  and Bioconductor}, pages 397--420. Springer, 2005.

\bibitem{amar2013dissection}
David Amar, Hershel Safer, and Ron Shamir.
\newblock Dissection of regulatory networks that are altered in disease via
  differential co-expression.
\newblock {\em PLoS Comput Biol}, 9(3):e1002955, 2013.

\bibitem{gambardella2013differential}
Gennaro Gambardella, Maria~Nicoletta Moretti, Rossella de~Cegli, Luca Cardone,
  Adriano Peron, and Diego di~Bernardo.
\newblock Differential network analysis for the identification of
  condition-specific pathway activity and regulation.
\newblock {\em Bioinformatics}, 29(14):1776--1785, 2013.

\bibitem{kuijjer2015estimating}
Marieke~Lydia Kuijjer, Matthew Tung, GuoCheng Yuan, John Quackenbush, and
  Kimberly Glass.
\newblock Estimating sample-specific regulatory networks.
\newblock {\em arXiv preprint arXiv:1505.06440}, 2015.

\bibitem{huang2008systematic}
Da~Wei Huang, Brad~T Sherman, and Richard~A Lempicki.
\newblock Systematic and integrative analysis of large gene lists using david
  bioinformatics resources.
\newblock {\em Nature protocols}, 4(1):44--57, 2008.

\bibitem{fresno2013rdavidwebservice}
Crist{\'o}bal Fresno and Elmer~A Fern{\'a}ndez.
\newblock Rdavidwebservice: a versatile r interface to david.
\newblock {\em Bioinformatics}, page btt487, 2013.

\bibitem{kalinka2011linkcomm}
Alex~T Kalinka and Pavel Tomancak.
\newblock linkcomm: an r package for the generation, visualization, and
  analysis of link communities in networks of arbitrary size and type.
\newblock {\em Bioinformatics}, 27(14), 2011.

\end{thebibliography}
\end{document}